\documentstyle[twocolumn,epsf,prl,aps]{revtex}

\newcommand {\be}{\begin{equation}}
\newcommand {\ee}{\end{equation}}
\newcommand {\bea}{\begin{eqnarray}}
\newcommand {\eea}{\end{eqnarray}}
\begin{document}

\draft

\title
{New Software for Measurements of the Anisotropic Resistivity by Multiterminal Technique} 
\author{ G. A. Levin\cite{X} and D. A. Chernikov }
\address{Virtual Instruments, 150 East 39th Street, Suite  804, New York,  NY 10016\\
\date{\today}
\vspace{.5cm}
{\rm \begin{quote}
We present an overview of the capabilities of a new software developed for use with multiterminal measurements. 
The program supports two configurations of current and voltage contacts   
known in the parlance of the field as "flux transformer" and " I$\|$c" geometries. 
The problem of instability of the results obtained by the multiterminal method and the ways to overcome it is also
discussed.
\end{quote}
}
}
\maketitle



\section{Introduction}

Even today, the most commonly used method of measuring resistivity is a 
four-point method in which a uniform distribution  of current is created along a certain 
direction in a sample of a given cross-section. This allows to determine one component
of the resistivity tensor.  The multiterminal method, in contrast, is based on creating a strongly
nonuniform two-dimensional distribution of current in a sample of a given geometry  and measuring
the voltage drops  between the two pairs of voltage contacts attached to the surface in different
places. Using the mathematical model of the current distribution, one calculates then  the two
components of the resistivity tensor from the measured voltages. 

The advantages of  the  multiterminal technique are obvious. 
One needs only one sample, instead of two, in order to measure two components of the 
resistivity of an anisotropic medium.  This alone reduces the time and expense of  an experiment in
half. Moreover, in many practically important cases the four-point method is difficult to apply due to
specific shape and small size of available samples. This has been the case with the crystals of high
critical temperature (high-$T_c$) superconductors, manganites exhibiting colossal magnetoresistance (CMR) 
and the number of other layered crystals available mostly in the form of  very thin platelets. 
This makes it difficult to measure the  resistivity component in the direction of the smallest
dimension. 

Perhaps even more important is that in some of the most interesting materials  
it is necessary to measure both components of the resistivity tensor on the same sample. Even the
crystals from the same batch may have somewhat different concentration of dopants or/and  impurities.
If the  resistivity strongly changes with elemental composition,  one cannot reliably compare the
temperature or magnetic field dependence of the  two components of the resistivity 
tensor if they are measured on different samples. 

Yet, the acceptance of the multiterminal technique for measurements of the
{\it normal state resistivities} has been slow even for research purposes, not to mention the engineering
applications.  Friedmann et al.\cite{Friedmann} have used a modified Montgomery method to measure 
the components of the resistivity tensor in untwinned single crystals of 
$YBa_2Cu_3O_{7-\delta}$.  Later, Ando et al.\cite{Ando} adopted an algorithm of Busch et
al.\cite{Busch} to obtain  the two components of the normal state resistivity ($\rho_c$ and
$\rho_{ab}$ ) of $Bi_2Sr_2CuO_y$  crystals in high magnetic field. 

A systematic approach to the development of the mathematical models 
describing two configurations of the current and voltage contacts, known in the parlance of the field
as  "flux transformer"  and "I$\|$c"  geometries,  is described in Refs.\cite{Jiang,Levin,LevinC}.
The numerical algorithms based on these models were used to measure  the resistivities of 
$PrBa_2Cu_3O_{7-\delta}$\cite{LevinPRL}, underdoped  $YBa_2Cu_3O_{7-\delta}$\cite{AlmasanJLT},
and electron-doped $Nd_{2-x}Ce_xCuO_{4-y}$\cite{AlmasanIJMP}. Two other research groups later have
developed  similar programs.   At  Argonne  the resistivities of CMR  manganites
have been measured by the multiterminal method\cite{Gray,Gray1}, and  an ISTEC group
have investigated the applicability of the multiterminal technique 
to the mixed state of superconductors\cite{Nakao}.

Apparently, the main obstacle  to a wider acceptance  of the multiterminal method 
has been  a relative complexity of the calculations involved in determining the values of the resistivities from the
measured voltages.   Here we present a brief outline of a new software  developed 
specifically for that purpose. One of the main goals   of this project has been to facilitate the
implementation of  the multiterminal technique 
in common practice by allowing other research groups to avoid  wasteful and costly duplication
of the development of the mathematical models,   programming  and testing so characteristic of the current
situation.  
\section{Software characteristics.}
The software package called {\it Ariadne} is available for download from the website \underline
{\bf www.virtinst.com}. The algorithm by which this program
calculates the components of the resistivity tensor is based on the method of Refs.\cite{Levin,LevinC}. 
However, the capabilities of our software are significantly enhanced 
by inclusion in the mathematical models of the flux transformer and I$\|$c  configurations the
options of  an arbitrary offset  of the current contacts from the edge of the sample and an arbitrary width of
the current and voltage contacts. This makes the models much more realistic and allows to carry out
a very important test of the stability of the results with respect to  variations of the input parameters.

The algorithms are thoroughly optimized. Even when
the current contacts are offset from the edges of the sample and have a finite width, it is possible to eliminate 
unnecessary summation of the slowly convergent series. As the result,  it takes Ariadne a fraction 
of a second to find a solution on a regular PC with the accuracy  of
$10^{-7}$ or better (needless to say that such an accuracy is  excessive given a much greater  error with
which  the geometrical parameters of the sample are typically measured). In our tests the  large data files with hundred
entries  have been processed in a few seconds.\\   
Other features include:\\
$\bullet$ The ability to run batch jobs, i.e. to process several input data files in a row without user interaction.\\
$\bullet$ Multithreaded processing, so that a user  has complete access to the user interface even while a job is
being run.\\
$\bullet$  Control over how the input files are formatted, output files are created and named, and what goes into
them.\\
$\bullet$ Comprehensive error reporting mechanism to ensure reliability of the calculations.\\
$\bullet$ Clean and intuitive interface.\\
$\bullet$ Context-sensitive help. 
\section{Stability of the Results.}
In the four-point method an error in determining the exact distance between the voltage contacts
results in equivalent relative error in the value of resistivity. 
The situation is more complex with multiterminal measurements. 
In certain cases a relatively small variations of the input parameteres result in much greater 
relative variations of at least one of the components of the resistivity tensor. 

One example of such 
"instability" is when the flux transformer configuration is used to determine the resistivities
of  a sample which is either too thin  or not anisotropic enough, so that the distribution of the density of current is 
close to uniform. The "vertical" component of the current density is small in comparison with the 
"horisontal" one and therefore the relative error in determining the respective component of the resistivity 
($\rho_c$ ) is necessarily   substantially greater than that for $\rho_{ab}$. 
However,  such instability has subtle  manifestations  and may remain undetected. 
Any software calculates the resistivities from the measured voltages with a  preset accuracy. 
This, however,  does not mean that the resistivities are indeed determined  with that accuracy.  
To detect the low stability of
the metod one has to calculate the resistivities from the same voltages using  slightly different values of the
geometrical parameters, especially those that are determined least precisely. In our example one would find that the
calculated values of $\rho_{ab}$ change very little with  the small variations of the width of the contacts and 
the value of their
offset from the edges of the samples. In contrast, the calculated values of $\rho_c$ vary much more strongly in
response to the variations  of these  input parameters.  

The solution of such a problem is to use either the alternative configuration (I$\|c$) 
or a combination of both, and to take as the final results the values of  $\rho_c$ and $\rho_{ab}$
that show the greatest stability with respect to  error in determination of the geometrical
parameters. Fast software like Ariadne which also allows to run batch jobs makes this tedious work 
much easier.

 

\end{document}